\documentclass[aps,pra,nobibnotes,twocolumn,superscriptaddress]{revtex4-1}
\usepackage{graphicx}
\usepackage{mathrsfs}
\usepackage{bm}
\usepackage{amsmath}
\usepackage{dcolumn}
\usepackage{epstopdf}
\usepackage{dsfont}
\usepackage{amssymb}
\usepackage{tabularx}
\usepackage{array}
\usepackage{float}
\usepackage{color}
\usepackage{epstopdf}
\usepackage{mathrsfs}
\usepackage[colorlinks, linkcolor=blue,anchorcolor=blue,citecolor=blue,urlcolor=blue]{hyperref}

\begin{document}

\preprint{APS/123-QED}

\title{Topological bosonic states on ribbons of honeycomb lattice}

\author{Yiping Wang}
\affiliation{Department of Physics, Key Laboratory of Micro-Nano Measurement-Manipulation and Physics (Ministry of Education), Beihang University, Beijing, 100191, China}

\author{Xingchuan Zhu}
\affiliation{Department of Physics, Beijing Normal University, Beijing, 100875, China}

\author{Kefei Zou}
\affiliation{Department of Physics, Key Laboratory of Micro-Nano Measurement-Manipulation and Physics (Ministry of Education), Beihang University, Beijing, 100191, China}

\author{Shengyuan A. Yang}
\affiliation{Research Laboratory for Quantum Materials, Singapore University of Technology and Design, Singapore 487372, Singapore}

\author{Huaiming Guo}
\email{hmguo@buaa.edu.cn}
\affiliation{Department of Physics, Key Laboratory of Micro-Nano Measurement-Manipulation and Physics (Ministry of Education), Beihang University, Beijing, 100191, China}


\begin{abstract}
The topological properties of hardcore bosons on ribbons of honeycomb lattice are studied using quantum Monte Carlo simulations. We map out a rich phase diagram with the superfluid and insulator phases at various fillings. Particularly, it is revealed that the insulator state at half filling is a topological bosonic state, which is characterized by a nontrivial Berry phase and a pair of edge states. We provide intuitive picture to understand this topological bosonic insulator state by showing that it can be adiabatically mapped to a topological fermionic model. It is also shown that the topological edge states are robust against weak interactions beyond the hardcore repulsion. Our results can be simulated by using bosonic cold atoms trapped in designed optical lattices.
\end{abstract}

\maketitle

\section{Introduction}
The study of topological phases has been at the frontier of current physics research~\cite{hasan2010,qi2011,bansil2016,haldane2017}. Over the past decade, much progress has been made in understanding and realizing various kinds of topological phases in various systems, particularly for Bloch electrons in crystalline materials. Based on commonly encountered anti-unitary symmetries, the noninteracting fermions can be classified into ten symmetry classes, and topologically nontrivial states can appear at different spatial dimensions~\cite{schnyder2008,chiu2016,Wen2017}. Further combined with lattice symmetries, much more symmetry-protected topological phases have been proposed and are actively searched for. However, the current experimental study of fermion topological phases lags behind the theoretical progress. The big challenges come from the material side: (i) the naturally existing topological materials are rare; (ii) topological band features are often complicated by the presence of other trivial extraneous bands in the same energy window; (iii) the study is further baffled by the lack of good control of the material properties, such as doping level, surface condition, defects, etc.

While most studies are based on fermions, there are increasing efforts to extend the topological properties to bosonic systems. From the experimental perspective, technqiues for realizing precisely controllable bosonic models have been well developed. For example, using ultracold atoms in optical lattices, one can in principle simulate any bosonic lattice models with arbitrary interactions. The optical/mechanical metamaterials also offer versatile playground for realizing various topological photonic/acoustic states. In addition, compared with fermions, bosonic systems also have their unique characters, e.g., bosons tend to condense, such that topological bosonic states only exist in interacting systems~\cite{he2015,Sterdyniak2015}; and their symmetry properties under time reversal or rotation are distinct from fermions. These reasons suggest a wealth of interesting physics to be explored in topological bosonic systems. Indeed, recent theoretical works have established rich topological phases for bosonic systems from studying the   cohomology classification and the phenomenological Chern-Simons field theory~\cite{chen2011,lu2012}. Still it is highly desirable to have simple and experimentally relevant topological bosonic models, which can allow transparent physical pictures that can be verified in experiment.

A natural idea to construct such models is to simply load bosons in the hardcore limit to the known topological lattice models for fermions. The allowed topological phases are likely to persist at a finite interaction.  In this way, a topological Bose-Mott insulator in a one-dimensional (1D) optical superlattice has been identified~\cite{zhu2013,grusdt2013,deng2014,Matsuda2014}. However this approach often fails for higher dimensions, or even quasi-1D systems (such as in the ladder geometry)~\cite{varney2010,guo2012}. The reason is that in strictly 1D open lattices, hardcore bosons behaves exactly the same as fermions due to the absence of particle exchange process, however, the commutation statistics of bosons breaks the topological phase when the exchange is possible for dimensions beyond strictly 1D. Nevertheless, the exchange processes as well as interactions opens opportunities to realize other nontrivial quantum phases~\cite{Takayoshi2013,Tovmasyan2013,vasi2015,greschner2016,Petrescu2017,greschner2018}, which is an interesting problem to explore.

As a prominent example of quasi-1D fermionic systems, the topological properties of graphene nanoribbons (GNRs) have been investigated by several works in the past\cite{gnr1,gnr2}. A recent work by Cao \emph{et al.} have shown that narrow GNRs with specific widths and termination realize 1D topological phases with protected end states~\cite{louie2017}. Specifically, for armchair GNRs, the spectrum is gapped for the width $N\neq 3n+2$ $(n>1)$. For such gapped GNRs (with a specific termination), topologically trivial and nontrivial phases appear in an oscillatory manner with increasing $N$. Moreover, the topology of a GNR can be modified by dopants or external fields. Thus, ribbons of honeycomb lattice provide an ideal model system of 1D topological phases. Naturally, it is interesting to study the physical properties of bosons in this geometry, which is readily accessible in current cold-atom experiments~\cite{Jotzu2014}. Since such ribbons are quasi-1D systems allowing boson exchange processes, the interplay between particle exchange, interaction, and lattice geometry may generate interesting topological phases.

\begin{figure}[htbp]
\centering \includegraphics[width=8.5cm]{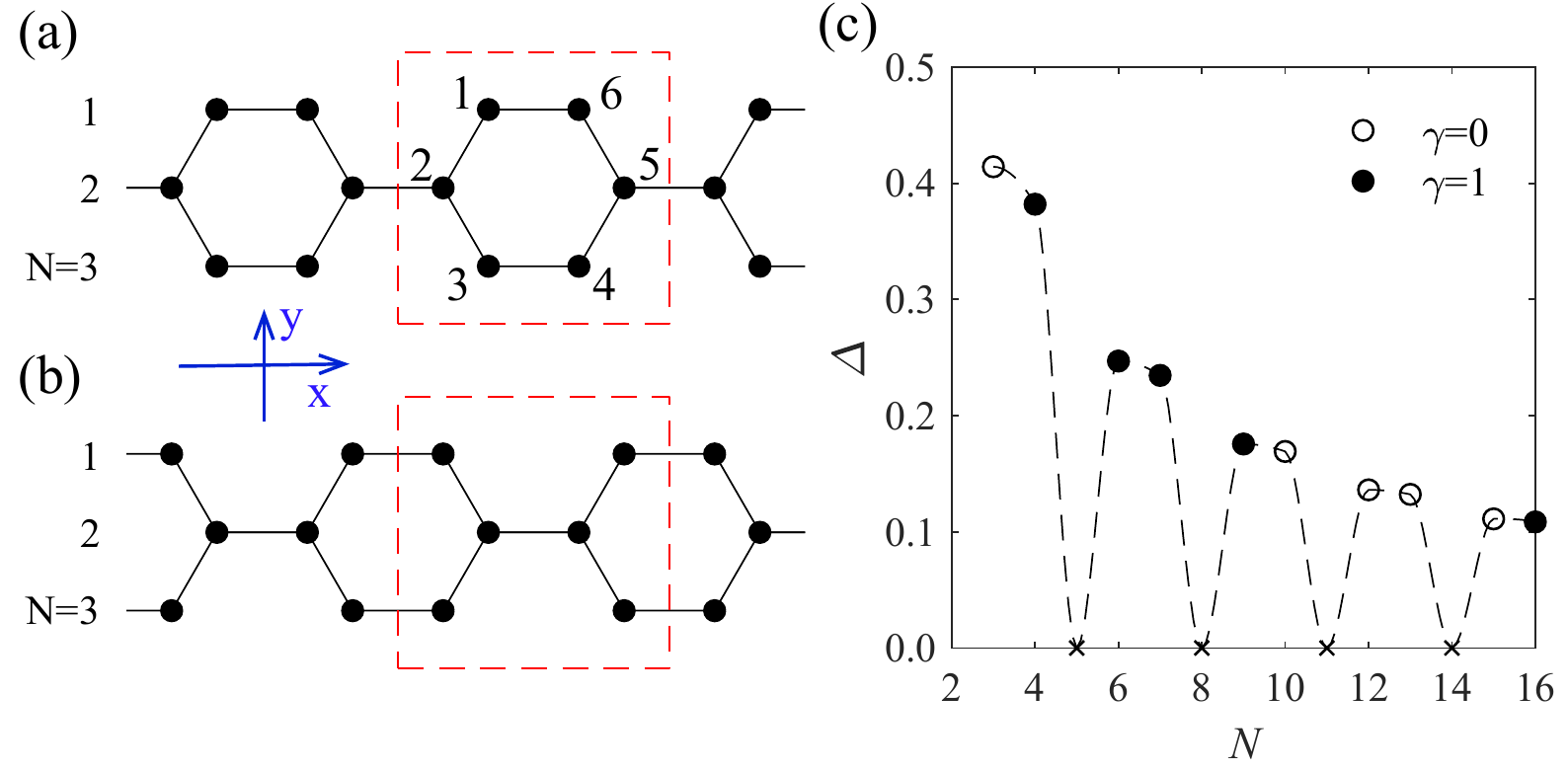} \caption{(a) and (b): schematics of a $N=3$ ribbon of honeycomb lattice with different unit cells. Each unit cell (see the region with red dashed boundary lines) contains six sites labeled from $1$ to $6$. Under open boundary condition, (a) [(b)] has termination A (B). Free fermions on geometry (a) form trivial insulator, while they have nontrivial topological property on geometry (b). (c): The gap of free fermions on the ribbon geometry as a function of the width $N$. The Berry phase for (a) is also shown using filled ($\gamma=1$) or empty ($\gamma=0$) circles.}
\label{fig1}
\end{figure}

Motivated by the recent progress mentioned above, in the paper, we study the topological properties of hardcore bosons on ribbons of honeycomb lattice using quantum Monte Carlo simulations. Through calculating the average density and superfluid density, we map out the phase diagram, which contains superfluid and insulator phases at various fillings. The nature of the insulators is analyzed. Interestingly, it is found that the insulator at half filling is a topological bosonic state, which is characterized by a nontrivial Berry phase and a pair of end states. We show that the topological bosonic insulator state is adiabatically connect to a limiting case that can be well understood using a related fermionic model. We also show that the topological bosonic end states are robust against weak interactions beyond the hardcore repulsion.

\section{model and approach}
We consider hardcore bosons loaded into armchair ribbons of honeycomb lattice. Hardcore means that the bosons have an infinitely large on-site repulsion, which forbids double occupancy of a single site. This behavior can be realized in ultracold atom experiment using Feshbach resonance \cite{feshbach}. The lattice geometry is depicted in Fig.\ref{fig1}. Here, the width $N$ of the ribbon counts the number of layers in the $y$ direction, as indicated in Fig.\ref{fig1}. The basic physics of this bosonic system is described by the following extended Bose-Hubbard model
\begin{eqnarray}\label{eq1}
H=-t\sum_{\langle i,j\rangle} (b_i^{\dagger}b_{j}+\text{H.c.})+\sum_{i}V_{i} n_i-\mu \sum_i n_i,
\end{eqnarray}
where $b_i$ ($b_i^{\dagger}$) is the hardcore bosonic annihilation (creation) operator, $n_i=b_i^{\dagger}b_i$ is the number operator for bosons. The occupying number of hardcore bosons is $0$ or $1$ on each site. Hence, the hardcore bosons obey commutation relation $[b_i,b_j^{\dagger}]=0$ for sites $i\neq j$ but anticommutation relation $\{b_i,b_i^{\dagger}\}=1$ for a single site $i$. This hardcore condition makes the model a strongly interacting one.  The first term in Eq.~(\ref{eq1}) is the nearest-neighbor hopping term, and the hopping amplitude $t$ will be taken as the unit of energy $(t=1)$ in our calculation. The second term in Eq.~(\ref{eq1}) represents a possible on-site potential. In the following, we take $V_i=V_0\cdot (y_i-\lfloor\frac{N}{2}\rfloor-1)$ where $y_i$ is the $y$ coordinate for the site $i$ (see Fig.\ref{fig1}), and $V_0$ the strength of the potential. Such a potential resembles the electrostatic potential for a transverse $E$ field in the case of GNRs, which is capable to drive a topological phase transition for the fermionic case. Here, we shall also investigate the possible phase transition driven by the potential $V$. Finally, $\mu$ in the last term denotes the chemical potential, which controls the number of bosons in the system.

The model in Eq.~(\ref{eq1}) has a $U(1)$ symmetry, namely the model is invariant under the transformation $b_i\rightarrow e^{i\theta}b_i$ where $\theta$ is a real-valued phase. In a superfluid phase, this symmetry would be spontaneously broken. The model also has a particle-hole symmetry at $\mu=0$ for odd $N$. For $\mu \neq0$, the average density $\rho_{\mu}$ at $\mu$ equals to the hole density $1-\rho_{-\mu}$ at $-\mu$, which makes the density versus $\mu$ curve centrosymmetric about the point $(0,0.5)$ in the $(\rho, \mu)$ plane. The system also respects a mirror symmetry $M_x:(x,y)\rightarrow(-x,y)$, which plays an important role in quantizing the Berry phase for the system, as we discuss below.
Moreover, we mention that the model is equivalent to a spin$-1/2$ $XY$ model through a mapping $S^{+}_i=b^{\dag}_i$ and $S^z_i=n_i-\frac{1}{2}$.

In the following, we employ the approach of stochastic series expansion (SSE) quantum Monte Carlo (QMC) method with directed loop updates~\cite{sandvik2002} to study the model in Eq.(1). The SSE method expands the partition function in power series and the trace is written as a sum of diagonal matrix elements. The directed loop updates make the simulation very efficient. Our simulations are on finite lattices with the total number of sites $N_s=6L$ for $L$ (the number of unit cells) up to $40$. There are no approximations
causing systematic errors, and the discrete configuration space can be sampled without floating
point operations. The temperature is set to be low enough to obtain the ground-state properties. For such quasi-1D bosonic systems, the notorious sign problem in the QMC approach can be avoided. In the following, we mainly focus on the $N=3$ case as shown in Fig.\ref{fig1}, and the results for other cases with larger $N$ are qualitatively similar.

\begin{figure}[htbp]
\centering \includegraphics[width=7cm]{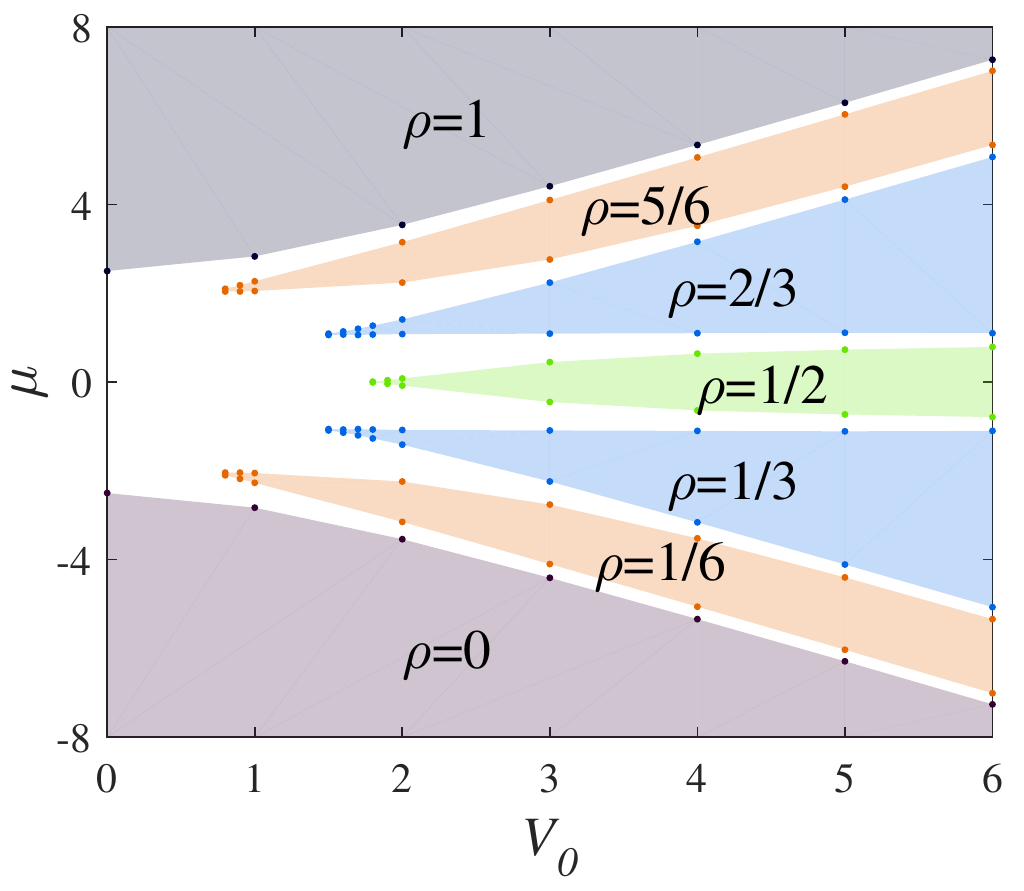} \caption{The phase diagram in $(\mu, V_0)$ plane, which contains superfluid and insulators at various fillings.}
\label{fig2}
\end{figure}

\section{phase diagram}
The phase diagram can be firstly understood in the atomic limit, when the hopping processes are turned off (by setting $t=0$). Then the occupancy of each site is determined by its on-site potential energy $V_i-\mu$. If $V_i-\mu<0$, a hardcore boson would be added to the site. For the $N=3$ case studied here, a $\rho=\frac{1}{3}$ insulator forms when $0>\mu>-V_0$; and a $\rho=\frac{2}{3}$ insulator forms when $V_0>\mu>0$.

When hopping is turned on, we find that naturally the range of the chemical potential for the above atomic insulators decreases and completely disappear at a critical value of $t/V_0$ (which depends on the specific insulator state).
The phase diagram obtained from QMC simulations is shown in Fig.~\ref{fig2}. One observes that the $\rho=\frac{1}{3}$ and $\rho=\frac{2}{3}$ atomic insulators persist at large $V_0$. Besides, there also appear insulators at the fillings $\rho=\frac{1}{6}, \frac{1}{2}, \frac{5}{6}$. These insulators are separated by incommensurate superfluid regions.

\begin{figure}[t!]
\centering \includegraphics[width=8cm]{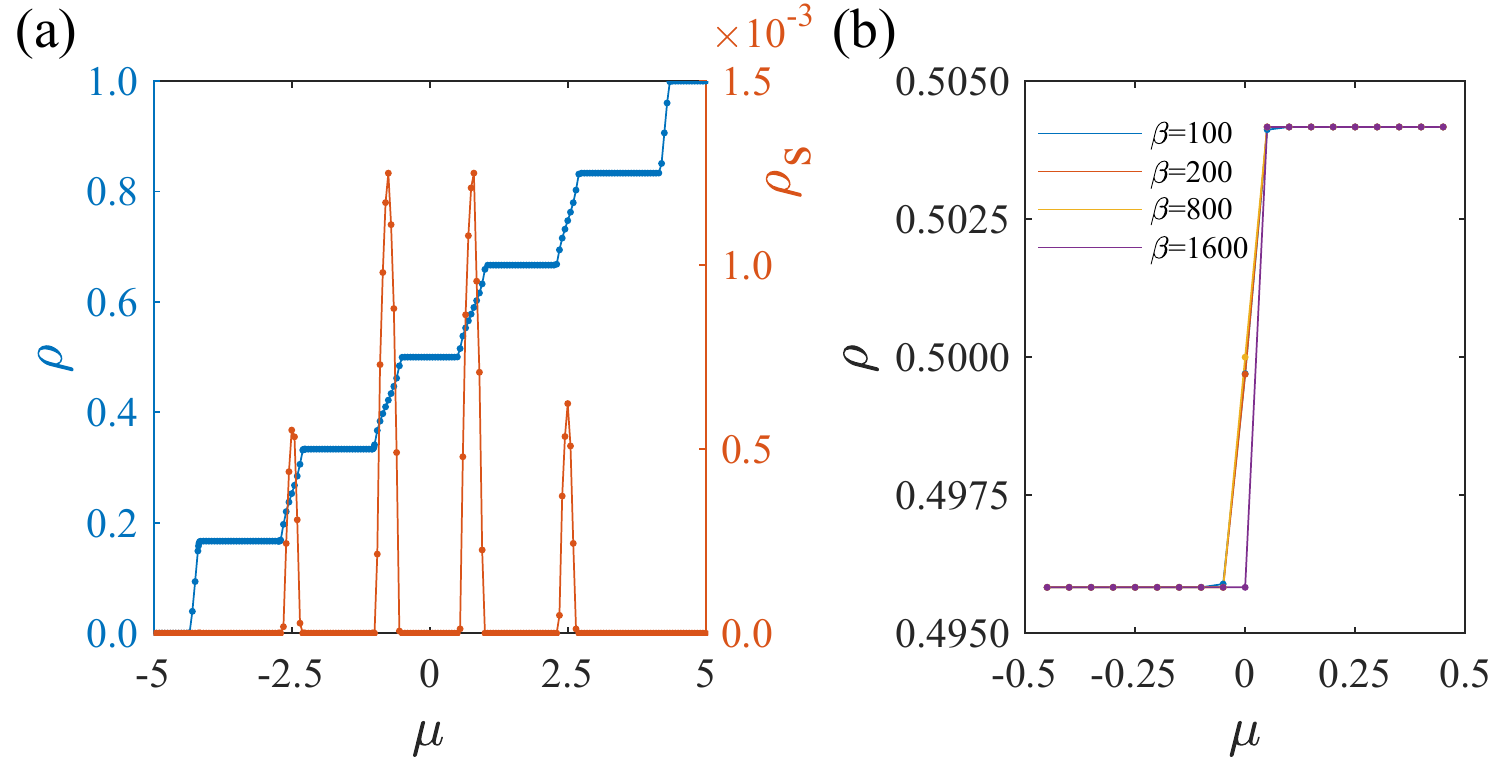} \caption{(a): The average density and superfluid density as a function of $\mu$ for $V_0=3$. (b): The splitting of the $\rho=0.5$ plateau at different temperatures under open boundary condition, which tends to be vertical in the limit $\beta\rightarrow\infty (T=0)$.}
\label{fig3}
\end{figure}

The phase diagram in Fig.~\ref{fig2} is obtained by explicitly calculating the average density $\rho=\frac{\sum_i n_i}{N_s}$ and the superfluid density $\rho_s$ in the QMC simulation. Here, the superfluid density is evaluated using the standard formula\cite{Li2015,Batrouni2006}
\begin{eqnarray}\label{eq2}
\rho_s=\frac{W^2}{2\beta t},
\end{eqnarray}
where $W$ is the winding number and $\beta$ the inverse temperature. An insulator is characterized by a plateaus in the density $\rho$ and by $\rho_s=0$. As shown in Fig.~\ref{fig3}, the average density exhibits a series of plateaus at commensurate fillings, on which the superfluid density vanishes. So these plateaus correspond to the incompressible insulator phases, whose gaps are given by the widths of the plateaus. Between the insulators, the average density increases continuously with the chemical potential. For small $V_0$, the superfluid density is finite implying the system is in a superfluid phase. For large $V_0$, the superfluid density with $\rho<\frac{1}{6}$ and $\rho>\frac{5}{6}$ is zero (see Fig.~\ref{fig3} where $V_0=3$). This can be understood by noticing that the large $V_0$ isolates the sites $3$ and $4$ in the unit cell, which form a dimer. So for $\rho<\frac{1}{6}$, part of the dimers are occupied and the system becomes a dimer insulator.

The $\rho=\frac{1}{6}, \frac{1}{2}, \frac{5}{6}$ insulator states are interesting, because they have no corresponding atomic limits. We now explore their nature. We calculate the local densities with periodic boundary condition using QMC. The local density varies inside the unit cell (along $y$ axis), but is uniform along the 1D ribbon. For the $\rho=\frac{1}{6}$ insulator, with $V_0>0$, one hardcore boson would mainly distribute on sites $3$ and $4$. The sites $3$ and $4$ can be thought as being largely isolated by the applied potential $V$, and form a dimer. This approximation becomes exact in the limit of $V_0\rightarrow \infty$. Thus the $\rho=\frac{1}{6}$ insulator is reminescent of the insulator comprising of isolated dimers.
With this scenery, the starting $\mu$ for the $\rho=\frac{1}{6}$ plateau can be estimated to be $-V_0-t$, which approaches the exact value $-t-\sqrt{V_0^2+2t^2}$ at large $V_0$ (see Appendix A for the derivation).
The $\rho=\frac{5}{6}$ insulator can be analyzed in a similar way using the hole representation.

The $\rho=\frac{1}{2}$ insulator is the most interesting. Using the similar argument as mentioned above, we see that this state is adiabatically connected to a chain insulator containing the sites $2,3,4$ and $5$ in the large $V_0$ limit. We shall look into this phase in more detail in the next section.

\section{Topological Mott insulator}

It has been shown that armchair GNRs can be topologically nontrivial, which depends on the type of termination. We consider two terminations $A$ and $B$, as sketched in Fig.\ref{fig1}. For the fermionic case, the applied electric field (which produces the potential $V_i$) can drive topological phase transitions. In the following, we investigate whether the analogous effect happens for our bosonic system at half filling.

\begin{figure}[htbp]
\centering \includegraphics[width=8cm]{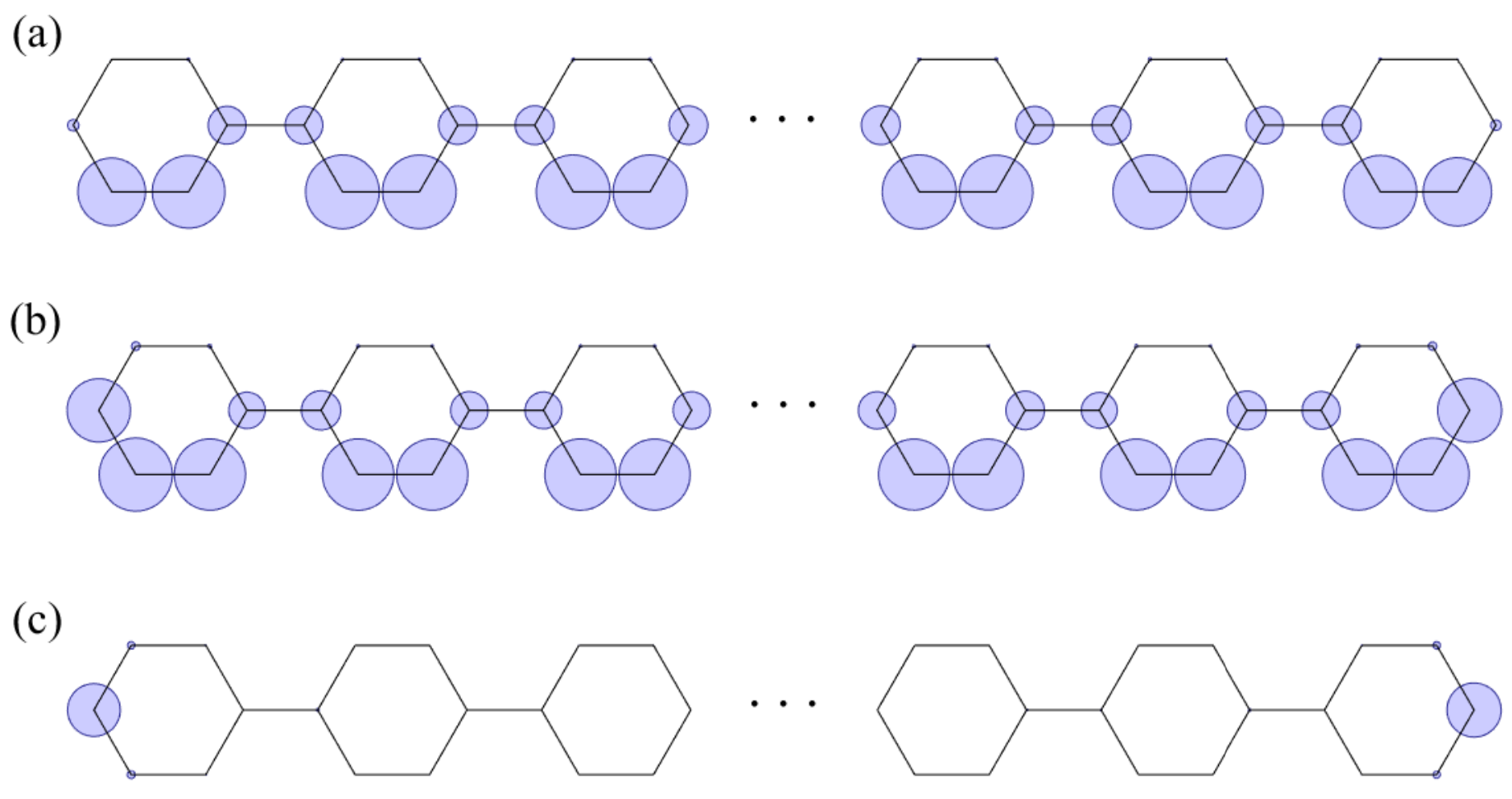} \caption{The local density under open boundary condition: (a) $\mu=-0.27$ approximately in the middle of the lower plateau; (b) $\mu=0.27$ approximately in the middle of the upper plateau. (c), The difference of the local densities in (a) and (b). Here $U_0=3$ and $\beta=100$.}
\label{fig4}
\end{figure}

Let's first investigate termination $A$.
The $\rho=\frac{1}{2}$ insulator appears at large $V_0$ is a bosonic Mott insulator. To check whether there exists nontrivial end states, we re-calculate the $\rho$-$\mu$ curve under the open boundary condition. In Fig.~\ref{fig3}(b), it shows that the plateau of the $\rho=\frac{1}{2}$ insulating phase is altered: the plateau splits into two pieces with a jump at a critical chemical potential in the middle. One of the split plateaus has $\rho_1=0.495833$, while the other has $\rho_2=0.504167$. Noticing that $N_s(\rho_2-\rho_1)=2$, the jump corresponds to the filling of two in-gap states, which can only be located at the two ends of the system, since the bulk is insulating.
Moreover, our calculation shows that the jump tends to be vertical in the limit of zero temperature.
These results imply that there appear two in-gap degenerate states for the open ribbon.

We further verify that these in-gap states are located at the two ends. The distributions of the harcore bosons under open boundary condition for two representative points on the split plateaus are calculated and plotted in Fig.~\ref{fig4}(a) and \ref{fig4}(b). One observes that the distribution of the bulk sites are nearly unchanged between the lower and the higher plateaus. The filling of the in-gap states happens only at the boundaries. When none of the in-gap states are filled (the lower plateau), there is $\frac{1}{2}$-fractional boson less at each boundary compared to the bulk sites. After both are filled (the higher plateau), there is $\frac{1}{2}$-fractional boson more. The difference in the local density between the upper and lower plateaus is shown in Fig.~\ref{fig4}(c).  It clearly demonstrates that the in-gap states are located near the ends. Thus our results provide clear evidence showing that the $\rho=\frac{1}{2}$ insulator is a topological bosonic Mott insulator. The above discussion is for termination $A$. In comparison, we do not observe the splitting of plateaus and end states for termination $B$, which indicates that it is a trivial insulator.

Topological states are generally characterized by topological invariants defined for the bulk. The $\rho=\frac{1}{2}$ topological Mott insulator here is characterized by a nontrivial Berry phase (or 1D winding number) defined for the many-body ground state with the twisted boundary phase~\cite{berry1,berry2,wang2014}:
\begin{eqnarray}\label{eq3}
 \gamma=i\oint \langle \psi_{\theta}|\frac{d}{d\theta}|\psi_{\theta}\rangle d\theta,
\end{eqnarray}
where $\theta$ varying from $0$ to $2\pi$ is the twisted boundary phase connecting the two ends of the system, and $\psi_{\theta}$ is ground-state wave-function corresponding to a particular $\theta$. We find that the Berry phase takes a nontrivial value of $\pi$ in the topological Mott insulator state. Note that for the current system, the Berry phase is quantized in unit of $\pi$ due to the presence of a mirror symmetry $M_x$. Thus, the insulator phase here is a symmetry-protected topological phase.
Here, the boundary termination affects the obtained Berry phase, because the different termination corresponds to different choice of the unit cell in the bulk (see Fig.\ref{fig1}). Our result shows that termination $A$ gives a nontrivial Berry phase, whereas termination $B$ gives a trivial Berry phase, which are consistent with the conclusion regarding the presence of end states.

As stated in the previous section, the topological Mott insulator state at half filling is adiabatically connected to the limiting case with $V_0$ going to infinity. It is intuitive and simple to understand the nontrivial topology in this limit. When $V_0\rightarrow\infty$, the occupied two rows with sites 2, 3, 4, and 5 can be viewed as isolated and forming a 1D chain. With Jordan-Wigner transformation $b^{\dagger}_{i}=c^{\dagger}_{i}e^{i\pi\sum^{i-1}_{k=1}c^{\dagger}_{k}c_{k}}$ ($c^{\dagger}_{i}$ the fermion creation operator),
the effective Hamiltonian can be mapped to a non-interacting fermionic one. For termination $A$, the obtained fermionic model reads
\begin{eqnarray}\label{eq3}
\widetilde{\mathcal{ H}}=-t\sum_{j}(c^{\dag}_j c_{j+1}+\textrm{H.c.})-V_0\sum_{i=4k+2,4k+3} c^{\dag}_{i}c_i,
 \end{eqnarray}
where $c_j$ and $c_j^\dagger$ are fermionic operators, and $k=0,1,2,...$ is an integer. Using a four-site unit cell, the Hamiltonian in the momentum space can be expressed as
\begin{eqnarray}\label{eq3}
\widetilde{\mathcal{ H}}(k_x)=\left[
               \begin{array}{cccc}
                 0 & -t & 0 & -te^{-ik_x} \\
                 -t & -V_0 & -t & 0 \\
                 0 & -t & -V_0 & -t \\
                 -te^{i k_x} & 0 & -t & 0 \\
               \end{array}
             \right].
\end{eqnarray}
Then the energy spectrum can be obtained directly and the gap at $\frac{3}{4}$ filling (corresponding to $\rho=\frac{1}{2}$ for the original lattice) is
\begin{eqnarray}\label{eq3}
\Delta=\sqrt{(U_0+2t)^2+4t^2}-\sqrt{(U_0-2t)^2+4t^2}.
\end{eqnarray}
The phase at $\frac{3}{4}$ filling is insulating for any $U_0>0$. By calculating the Berry phase, we find that the insulator is topological and a pair of degenerate in-gap states appear under the open boundary condition, consistent with the results from QMC.

\begin{figure}[htbp]
\centering \includegraphics[width=8.5cm]{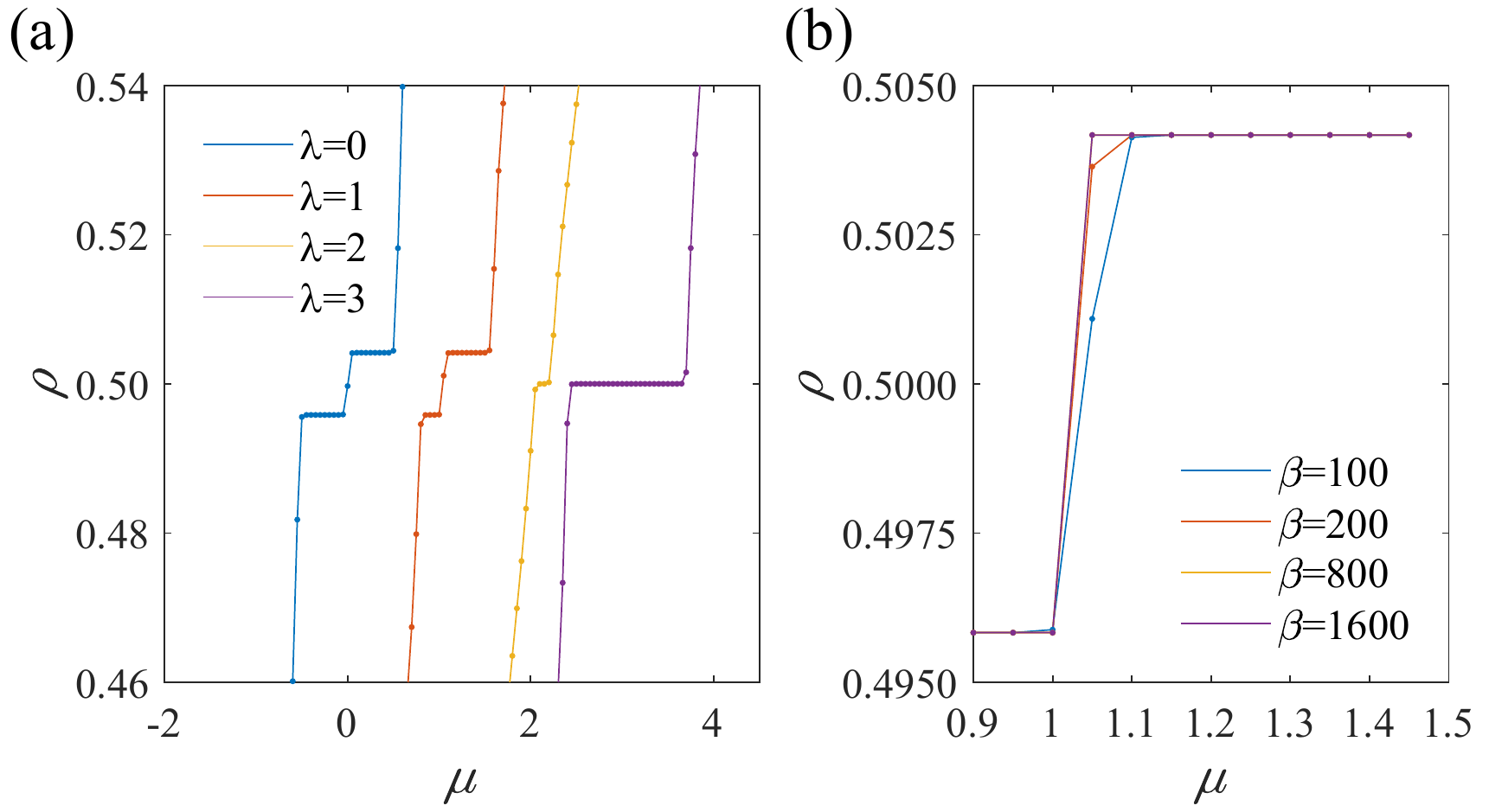} \caption{The average density as a function of $\mu$ under open boundary condition. (a): various $\lambda$ at fixed $\beta=100$. (b): different inverse temperatures at fixed $\lambda=1$.}
\label{fig5}
\end{figure}

\section{Effect of nearest-neighbor interaction}

So far, we have considered the hardcore boson interaction which is of on-site type. In practice, there may also be interactions between bosons at different sites. Whether these interactions would affect the stability of the topological Mott insulator phase is an interesting question to explore.
To study this, we include into our model Eq.~(\ref{eq1}) a nearest-neighbor interaction term
given by
\begin{eqnarray}\label{eq3}
H_\text{nn}=\lambda\sum_{\langle i,j\rangle}n_in_j,
\end{eqnarray}
with $\lambda$ the strength of the interaction.

We perform the QMC simulation with this interaction term added. Figure~\ref{fig5}(a) shows the average density as a function of $\mu$ under the open boundary condition with $\lambda=1,2,3$. When the interaction is relatively weak (e.g., $\lambda=1$), the pattern of split plateaus remains and the jump tends to be vertical in the $T=0$ limit [see Fig.\ref{fig5} (b)]. This demonstrates that the topological end states are robust against weak interactions. Nevertheless, one observes that the width of the lower plateau is shortened, implying that the gap size is reduced by the interaction. Increasing the interaction strength to $\lambda=2$, the pattern of split plateaus disappears. A new plateau at $\rho=0.5$ emerges and its width increases as the interaction is further increased. This result shows that a large enough interaction will completely destroy the nontrivial gap and a new insulator favored by the interaction would then be developed.

\section{Discussion and Conclusion}

In this work, we propose a simple quasi-1D topological bosonic model. Our model consists of hardcore bosons loaded onto a ribbon of honeycomb lattice. The properties of this model is studied using the quantum Monte Carlo approach. The superfluid and insulator phases at various fillings are identified and the phase diagram is mapped out. It is found that the insulator at half filling is a bosonic topological Mott insulator state, which is characterized by a nontrivial Berry phase and a pair of topological end states. We show that the bosonic topological Mott insulator is adiabatically connected to a limiting case that can be well understood using a related noninteracting fermionic model. It is also shown that the topological end states are robust against weak interactions beyond the hardcore repulsion.

Regarding the experimental study of our proposed model, we note that the experimental setup with cold atoms in optical lattice can be an ideal choice to realize the Bose-Hubbard model. Optical lattice with the honeycomb geometry has been demonstrated in experiment~\cite{Polini2013}. Moreover it is feasible to confine cold atoms to a specific region and subject them to an external potential like $V_i$ in our model. Hence, the experimental technique for realizing our model Eq.~(\ref{eq1}) is available. Furthermore, recently, the Berry phase for cold atoms in optical lattice was directly measured using a combination of Bloch oscillations and Ramsey interferometry~\cite{Atala2013}. Thus, the bosonic topological Mott insulator phase we discussed might be directly probed in experiment.

\section{Acknowledgments}
The authors thank Xuefeng Zhang for helpful discussion. H.G. acknowledges support from the NSFC grant No.~11774019. S.A.Y acknowledges support from the Singapore MOE AcRF Tier 2 (Grant No.~MOE2015-T2-2-144).

\appendix
\section{The band structure of the honeycomb ribbon}
The starting $\mu$ for $\rho\neq 0$ can be analytically determined by the band bottom of the honeycomb ribbon. One hardcore boson has exactly the same energy as one fermion due to the absence of exchanging statistics. The Hamiltonian in the momentum space writes as
\begin{eqnarray}\label{eqa1}
 H(k_x)=
\left[
  \begin{array}{cccccc}
    V_0 & -t & 0 & 0 & 0 & -t \\
    -t & 0 & -t & 0 & -te^{-ik_x} & 0 \\
    0 & -t & -V_0 & -t & 0 & 0 \\
    0 & 0 & -t & -V_0 & -t & 0 \\
    0 & -te^{ik_x} & 0 & -t & 0 & -t \\
    -t & 0 & 0 & 0 & -t & V_0 \\
  \end{array}
\right].
\end{eqnarray}
The energy spectrum contains six branches, which are $E_{1,2}=\pm t, E_{3,4}=\pm(t-\sqrt{2t^2+V^2_0}), E_{5,6}=\pm(t+\sqrt{2t^2+V^2_0})$ for $k_x=0$. The band bottom is located at $k_x=0$. The value of the band bottom $-(t+\sqrt{2t^2+V^2_0})$ gives the lower boundary of the $\rho=\frac{1}{6}$ region in the phase diagram.


\bibliography{ribbon_ref}

\end{document}